\documentstyle[12pt]{article}
\begin{document}
\setcounter{page}{1}
\thispagestyle{empty}

\centerline{\normalsize hep-ph/9406294 \hfill OITS 544}
\centerline{\normalsize \hfill June, 1994}

\vspace{15mm}

\begin{center} {\bf MINIMAL SUPERSYMMETRIC HIGGS BOSON DECAY RATE\\
                      IN $O(\alpha_{s}^{2})$ PERTURBATIVE QCD}\\

\vspace{1cm}
               {\bf Levan R. Surguladze} \\
\vspace{2mm}
{\it Institute of Theoretical Science, University of Oregon\\
                  Eugene, OR 97403, USA}

\vspace{3mm}
\end{center}
\begin{abstract}
A short presentation of the results of the analytical evaluation of
the $O(\alpha_s^2)$ QCD contributions in the fermionic decay rates
of the neutral CP-odd minimal supersymmetric Higgs particle is made.
The corrections due to the nonvanishing quark masses are included.
The results are presented both in terms of running and pole quark
masses.
\end{abstract}

\vspace{1cm}

    A crucial test of the most attractive extension of the Standard Model (SM)
- the Minimal Supersymmetric Standard Model (MSSM) can possibly be done at
LEP200. Indeed, as was shown in \cite{revSUSYLEP}, there is a significant
mass parameter space increase in the two Higgs doublet model at LEP200,
while for the SM Higgs the mass range will increase
only marginally. For the MSSM, a large theoretically allowed parameter
region can be covered \cite{revSUSYLEP} with $\sqrt{s}=210$ GeV,
highest expected luminosity of 500 pb$^{-1}$ and with the latest CDF result
for the top mass around 174 GeV \cite{CDF}. (For earlier analyses
of the MSSM Higgs discovery potential at LEP and LHC see \cite{Kun}
and references therein.)

  To study the detectability of various Higgs particles,
it is important  to start with the search of qualitative as well as
quantitive information about their decay processes.
One of the most interesting experimentally accessible quantities - the
branching fractions of the quark-antiquark decay modes can be
calculated within the perturbation theory. (For a review of the SM Higgs
phenomenology, taking into account radiative effects see \cite{rev1} and
references therein.)
A precise theoretical evaluation of decay rates is particularly important
from a viewpoint of distinction of MSSM and minimal SM Higgs signals.
In the recent work \cite{My} the fermionic decay rates of the minimal SM
Higgs particle have been evaluated to the $O(\alpha_s^2)$ in perturbative QCD.
In the present work, using the same technique, the $O(\alpha_s^2)$ QCD
contributions to the decay rates of the MSSM pseudoscalar Higgs particle
into the quark-antiquark pairs are evaluated analytically, taking into
account the nonvanishing quark masses. The assumption was made,
that all superpartners are too heavy to appear in the decays of the
Higgs boson.

   The minimal supersymmetric extension of the SM is the two-Higgs doublet
model. The physical Higgs boson spectrum of the MSSM consists of five states:
a charged Higgs pair $H^{\pm}$, neutral $CP$-even scalars $H_{s}$, $H'_{s}$
and a neutral $CP$-odd pseudoscalar $H_{p}$ (for a review see, e.g., the
textbook \cite{rev0}).
The Lagrangian density, describing the Yukawa interaction of the neutral
Higgs boson with quarks and leptons has the following form:
\begin{eqnarray}
\lefteqn{\hspace{-3cm}L =(\sqrt{2}G_F)^{1/2}\sum_{f}(C_{H_sff}(\alpha,\beta)
                        m_f\overline{q}_fq_fH_s
                       +C_{H_pff}(\beta)m_f\overline{q}_fi\gamma_5q_fH_p)}
                                                                 \nonumber\\
 && \quad \hspace{4cm}
             \equiv (\sqrt{2}G_F)^{1/2}\sum_{f}( j_f^{(s)}H_s+j_f^{(p)}H_p)
\label{eq:Lagr}
\end{eqnarray}
Below $q_f$ denote the quark fields with flavor $f$ and mass $m_f$.
The coefficients $C_{H_sff}(\alpha,\beta)$ and $C_{H_pff}(\beta)$
are the functions of
the ratio of the Higgs-field vacuum expectation values
- tan$\beta$ and a mixing angle $\alpha$ in the neutral $CP$-even sector.
The expressions for $C_{H_sff}(\alpha,\beta)$ and $C_{H_pff}(\beta)$
for the particular quark or lepton pair can be found, e.g., in \cite{rev0}.
In the SM limit $C_{H_sff}(\alpha,\beta)=-1$ and the pseudoscalar term
drops out in the eq.(\ref{eq:Lagr}). The corresponding calculation
of the $O(\alpha_s^2)$ SM Higgs decay rate (relevant for the MSSM Higgs
scalar as well) has been done in \cite{My}.
In the present work one considers the pseudoscalar Higgs, expecting
that the reader is familiar with the ref.\cite{My}.

        The two-point correlation function of the pseudoscalar currents
$j_f^{(p)}$ has the following form:
\begin{equation}
\Pi(Q^2=-s,m_f)=i\int e^{iqx}<Tj_f^{(p)}(x)j_f^{(p)}(0)>_{0}d^4x.
\label{eq:PI}
\end{equation}
For the decay width one has:
\begin{equation}
\Gamma_{H_{p}\rightarrow q_f\overline{q}_f}
         =\frac{\sqrt{2}G_F}{M_{H_p}} Im\Pi(s+i0,m_f)\biggr|_{s=M_{H_p}^2}.
\label{eq:ImPi}
\end{equation}

    The full $O(\alpha_s)$ analytical result
for the decay rate of $H_{p}\rightarrow q_f\overline{q}_f$
in terms of pole quark masses looks like \cite{Dre}:
\begin{equation}
\Gamma_{H_{p}\rightarrow q_f\overline{q}_f}
      =\frac{3\sqrt{2}G_FM_{H_p}}{8\pi}C_{H_{p}ff}(\beta)m_f^2
           \biggl(1-\frac{4m_f^2}{M_{H_p}^2}\biggr)^{\frac{1}{2}}
             \biggl[1+\frac{\alpha_s(M_{H_p})}{\pi}
                       \delta^{(1)}(\frac{m_f^2}{M_{H_p}^2})
               +O(\alpha_s^2)\biggr],
\label{eq:2loop}
\end{equation}
where:
\begin{displaymath}
\delta^{(1)}=\frac{4}{3}\biggl[\frac{a(\eta)}{\eta}
     +\frac{19+2\eta^2+3\eta^4}{16\eta}\log\gamma
         +\frac{21-3\eta^2}{8}\biggr],
\end{displaymath}
\begin{displaymath}
a(\eta)=(1+\eta^2)\biggl[4Li_2(\gamma^{-1})+2Li_2(-\gamma^{-1})
          -\log\gamma\log\frac{8\eta^2}{(1+\eta)^3}\biggr]
           -\eta\log\frac{64\eta^4}{(1-\eta^2)^3},
\end{displaymath}
\begin{displaymath}
\gamma=\frac{1+\eta}{1-\eta},
\hspace{5mm}
\eta=\biggl(1-\frac{4m_f^2}{M_{H_p}^2}\biggr)^{\frac{1}{2}}.
\end{displaymath}

    The expansion of the r.h.s of eq.(\ref{eq:2loop}) in a power series
in terms of small $m_f^2/M_{H_p}^2$ has the following form:
\begin{eqnarray}
\lefteqn{\Gamma_{H_{p}\rightarrow q_f\overline{q}_f}
                  =\frac{3\sqrt{2}G_FM_{H_p}}{8\pi}C_{H_{p}ff}(\beta)m_f^2
        \biggl\{\biggl(1-2\frac{m_f^2}{M_{H_p}^2}+...\biggr)} \nonumber\\
 && \quad
     \hspace{-5mm} +\frac{\alpha_s(M_{H_p})}{\pi}
               \biggl[3-2\log\frac{M_{H_p}^2}{m_f^2}
          +\frac{m_f^2}{M_{H_p}^2}\biggl(8+8\log\frac{M_{H_p}^2}{m_f^2}\biggr)
              +...\biggr]+O(\alpha_s^2)\biggr\},
\label{eq:2loopexpan}
\end{eqnarray}
where the period covers high order terms $\sim (m_f/M_{H_p})^{2k}$, $k=2,3...$.

The evaluation of the $O(\alpha_s^2)$ corrections to the eqs.
(\ref{eq:2loop},\ref{eq:2loopexpan}) has been done in a close analogy
of the similar calculation of the decay rate of SM Higgs boson
\cite{My}. Namely,
the full two-point correlation function (\ref{eq:PI}) was expanded in powers
of $m_f^2/Q^2$ in the ``deep'' Euclidean region ($Q^2 \gg m_f^2$).
The first two coefficient functions in this expansion was evaluated
to $O(\alpha_s^2)$ by applying an appropriate projectors to the
$\Pi(Q^2,m_f^B,m_{v}^B)$. $m_{v}$ is the mass of the quark appearing
virtually in some topological types of three-loop diagrams and ''B`` labels
the unrenormalized quantities.
In the calculation of divergent Feynman integrals
    the dimensional regularization formalism
\cite{drg} and the minimal subtraction prescription \cite{MS}
in its modified form - $\overline{MS}$ \cite{MSB} are used.
The $\gamma_5$ matrix is defined in $D$-dimensional space-time as an
object with the following properties:
\begin{displaymath}
\{\gamma_5,\gamma_{\mu}\}=0, \hspace{5mm} \gamma_5\gamma_5=1.
\end{displaymath}
The above definition causes no problems in dimensional regularization
when there are two $\gamma_5$ matrices in a closed fermionic loop.
The corresponding one-, two- and three-loop diagrams have been
evaluated analytically, using the special computer program HEPLoops
\cite{HEPL}.
The graph-by-graph results were summed up with
an appropriate symmetry and gauge group weights. In the obtained
expressions for the $\Pi_{i}$ in terms of bare quantities
one renormalizes the coupling and the quark mass in a standard way
(see, e.g., \cite{My}).
Finally, one should analytically continue the obtained $\Pi_{i}$
from Euclidean to Minkowski space and take the imaginary part at
$s=M_{H_p}^2$ ( eq.(\ref{eq:ImPi}) ). One obtains the
following analytical result for the standard QCD with $SU_c(3)$
gauge group:
\newpage
\begin{eqnarray}
\lefteqn{\Gamma_{H_p\rightarrow q_f\overline{q}_f}
                  =\frac{3\sqrt{2}G_FM_{H_p}}{8\pi}C_{H_{p}ff}(\beta)
        m_f^2\biggl\{
  1+\frac{\alpha_s}{\pi}\biggl(\frac{17}{3}
           +2\log\frac{\mu_{\overline{MS}}^2}{M_{H_p}^2}\biggr)}\nonumber\\
 && \quad
   +\biggl(\frac{\alpha_s}{\pi}\biggr)^2
      \biggl[\frac{10801}{144}-\frac{19}{2}\zeta(2)-\frac{39}{2}\zeta(3)
         +\frac{106}{3}\log\frac{\mu_{\overline{MS}}^2}{M_{H_p}^2}
         +\frac{19}{4}\log^2\frac{\mu_{\overline{MS}}^2}{M_{H_p}^2}\nonumber\\
 && \quad \hspace{1cm}
  -N\biggl(\frac{65}{24}-\frac{1}{3}\zeta(2)-\frac{2}{3}\zeta(3)
      +\frac{11}{9}\log\frac{\mu_{\overline{MS}}^2}{M_{H_p}^2}
      +\frac{1}{6}\log^2\frac{\mu_{\overline{MS}}^2}{M_{H_p}^2}\biggr)\biggr]
                                                               \nonumber\\
 && \quad \hspace{35mm}
     -\frac{m_f^2}{M_{H_p}^2}\biggl<
        2+\frac{\alpha_s}{\pi}\biggl(\frac{8}{3}
         +8\log\frac{\mu_{\overline{MS}}^2}{M_{H_p}^2}\biggr)\nonumber\\
 && \quad
   +\biggl(\frac{\alpha_s}{\pi}\biggr)^2
      \biggl[\frac{1429}{36}-54\zeta(2)-\frac{166}{3}\zeta(3)
         +\frac{155}{3}\log\frac{\mu_{\overline{MS}}^2}{M_{H_p}^2}
         +27\log^2\frac{\mu_{\overline{MS}}^2}{M_{H_p}^2}\nonumber\\
 && \quad \hspace{1cm}
  -N\biggl(\frac{3}{2}-\frac{4}{3}\zeta(2)-\frac{4}{3}\zeta(3)
     +\frac{14}{9}\log\frac{\mu_{\overline{MS}}^2}{M_{H_p}^2}
     +\frac{2}{3}\log^2
           \frac{\mu_{\overline{MS}}^2}{M_{H_p}^2}\biggr)\biggr]\biggr>
                                                               \nonumber\\
 && \quad
    +\biggl(\frac{\alpha_s}{\pi}\biggr)^2
     \sum_{v=u,d,s,c,b}\frac{m_{v}^2}{M_{H_p}^2}4 \biggr\},
\label{eq:gamma0}
\end{eqnarray}
where the Riemann functions $\zeta(2)=\pi^2/6$ and $\zeta(3)=1.202056903$.
The last term in eq.(\ref{eq:gamma0}) represents the contributions
from the three-loop diagrams containing the virtual quark loop
(see fig.1 of the ref.\cite{My}).
The  ``triangle anomaly'' type correction
(see fig.2 of the ref.\cite{My}) requires a special treatment because
of $\gamma_5$ in dimensional regularization, and is not included here.
However, it is reasonable to expect that their contribution will not
exceed 1\%. Note that the "triangle anomaly" type contributions vanish
identically in the zero quark mass limit. Note also the smallness of
corresponding corrections in the scalar channel \cite{My}.

     The leading three-loop term $\sim m_f^2$ (the massless approximation)
coincides with the one obtained in \cite{MPL,PRV,My}, while the three-loop
results $\sim m_f^4, m_f^2m_{v}^2$ are new.

   The calculated decay width obeys the homogeneous renormalization group
equation:
\begin{equation}
\biggl(\mu^2\frac{\partial}{\partial\mu^2}
   +\beta(\alpha_s)\alpha_s\frac{\partial}{\partial\alpha_s}
   -\gamma_m(\alpha_s)\sum_{l=f,v}
        m_l\frac{\partial}{\partial m_l} \biggr)
    \Gamma_{H_{p}\rightarrow q_f\overline{q}_f}
            \biggl(\frac{\mu^2}{M_{H_p}^2},m_f,m_{v},\alpha_s\biggr)=0.
\label{eq:RG}
\end{equation}
For the standard definition of the QCD $\beta$-function and the mass
anomalous dimension see the, e.g., ref.\cite{My} (eqs. (14)-(17)).
The renormalization group relates the coefficients of $\log$ terms in
eq.(\ref{eq:gamma0}). The corresponding relations up to $O(\alpha_s^3)$
for the SM Higgs decay, which are valid for the pseudoscalar Higgs as well,
has been obtained in \cite{My}.

   The solution of the renormalization group equation (\ref{eq:RG})
at $\mu^2_{\overline{MS}}=M_{H_p}^2$ has the following form:
\newpage
\begin{eqnarray}
\lefteqn{\hspace{-1cm}\Gamma_{H_p\rightarrow q_f\overline{q}_f}
     =\frac{3\sqrt{2}G_FM_{H_p}}{8\pi}C_{H_{p}ff}(\beta)
       m_f^2(M_{H_p})\biggl\{
                1-2\frac{m_f^2(M_{H_p})}{M_{H_p}^2}
  +\frac{\alpha_s(M_{H_p})}{\pi}
         \biggl(5.66667-\frac{8}{3}\frac{m_f^2(M_{H_p})}{M_{H_p}^2}\biggr)}
                                                             \nonumber\\
 && \quad \hspace{7mm}
  +\biggl(\frac{\alpha_s(M_{H_p})}{\pi}\biggr)^2
     \biggl[35.93996-1.35865N
        +\frac{m_f^2(M_{H_p})}{M_{H_p}^2}\biggl(115.64581-2.29599N\biggr)
                                                              \nonumber\\
 && \quad \hspace{3cm}
            +4\sum_{v=u,d,s,c,b}\frac{m_{v}^2(M_{H_p})}
                                     {M_{H_p}^2}\biggr]\biggr\}.
\label{eq:RGimpr}
\end{eqnarray}
For the standard parametrization of the running coupling and the running mass
see, e.g., ref.\cite{My}.
One can see that the three-loop mass correction is positive in contrary
to the analogous result for the SM Higgs \cite{My}.

The
relation between the $\overline{MS}$ quark mass $m_f(M)$ renormalized
at arbitrary $M$ and evaluated for the $N$-flavor theory and the pole quark
mass $m_f$ has the following form \cite{My}:
\begin{eqnarray}
\lefteqn{\hspace{-1cm} m_f^{(N)}(M)=m_f\biggl\{1
   -\frac{\alpha_s^{(N)}(M)}{\pi}
      \biggl(\frac{4}{3}+\gamma_0\log\frac{M^2}{m_f^2}\biggr)
      -\biggl(\frac{\alpha_s^{(N)}(M)}{\pi}\biggr)^2\biggl[K_f}
                                                                \nonumber\\
 && \quad
        +\sum_{m_f<m_{f'}<M}\delta(m_f,m_{f'})-\frac{16}{9}
        +\biggl(\gamma_1^{(N)}-\frac{4}{3}\gamma_0
          +\frac{4}{3}\beta_0^{(N)}\biggr)\log\frac{M^2}{m_f^2}
                                                             \nonumber\\
 && \quad \hspace{7cm}
      +\frac{\gamma_0}{2}(\beta_0^{(N)}-\gamma_0)\log^2\frac{M^2}{m_f^2}
                                                     \biggr]\biggr\},
\label{eq:mMHtopole}
\end{eqnarray}
where $\beta_0^{(N)}$, $\gamma_0$ and $\gamma_1^{(N)}$ are the perturbative
coefficients of the QCD $\beta$ function and the mass anomalous dimension
$\gamma_m$ correspondingly and can be found, e.g., in ref.\cite{My}.
$K_f$ and $\delta(m_f,m_{f'})$ can be obtained from the on-shell results
of \cite{BRH} (see \cite{My}):
\begin{equation}
K_f= \frac{3817}{288}+\frac{2}{3}(2+\log2)\zeta(2)-\frac{1}{6}\zeta(3)
  -\frac{N_f}{3}\biggl(\zeta(2)+\frac{71}{48}\biggr)
  +\frac{4}{3}\sum_{m_l \leq m_f} \Delta\biggl(\frac{m_l}{m_f}\biggr),
\label{eq:K}
\end{equation}
\begin{equation}
\delta(m_f,m_{f'})=-\frac{1}{3}\zeta(2)-\frac{71}{144}
    +\frac{4}{3}\Delta\biggl(\frac{m_{f'}}{m_f}\biggr),
\label{eq:deltaM}
\end{equation}
\begin{equation}
\Delta(r)=\frac{1}{4}\biggl[\log^2r+\zeta(2)-\biggl(\log r
        +\frac{3}{2}\biggr)r^2
        -(1+r)(1+r^3)L_{+}(r)-(1-r)(1-r^3)L_{-}(r)\biggr],
\label{eq:delta}
\end{equation}
\begin{displaymath}
L_{\pm}(r) = \int_{0}^{1/r}dx\frac{\log x}{x \pm 1}.
\end{displaymath}
$L_{\pm}(r)$ can be evaluated for different quark mass ratios
$r$ numerically (see the table 1 of the ref.\cite{My}).
In the above equations the number of participating quark flavors
$N$ is specified according to the size of $M$ and has
no correlation with the quark mass $m_f$.

    Substituting eqs. (\ref{eq:mMHtopole})-(\ref{eq:delta})
at $M=M_{H_p}$ and the coefficients of the renormalization group functions
into the eq.(\ref{eq:RGimpr}), one obtains  the general form for the
decay rate $\Gamma_{H_{p}\rightarrow q_f\overline{q}_f}$
in terms of the pole quark masses:
\begin{eqnarray}
\lefteqn{\hspace{-1cm}\Gamma_{H_p\rightarrow q_f\overline{q}_f}
     =\frac{3\sqrt{2}G_FM_{H_p}}{8\pi}C_{H_{p}ff}(\beta)
               m_f^2\biggl\{1-2\frac{m_f^2}{M_{H_p}^2}}
                                                       \nonumber\\
 && \quad \hspace{1cm}
  +\frac{\alpha_s^{(N)}(M_{H_p})}{\pi}
   \biggl[3-2\log\frac{M_{H_p}^2}{m_f^2}
   +\frac{m_f^2}{M_{H_p}^2}\biggl(8+8\log\frac{M_{H_p}^2}{m_f^2}\biggr)\biggr]
                                                             \nonumber\\
 && \quad \hspace{1cm}
  +\biggl(\frac{\alpha_s^{(N)}(M_{H_p})}{\pi}\biggr)^2
   \biggl<\frac{697}{18}-\biggl(\frac{73}{6}+\frac{4}{3}\log 2\biggr)\zeta(2)
          -\frac{115}{6}\zeta(3)
                                                             \nonumber\\
 && \quad \hspace{40mm}
          -N\biggl(\frac{31}{18}-\zeta(2)-\frac{2}{3}\zeta(3)\biggr)
                                                             \nonumber\\
 && \quad \hspace{5mm}
    +\frac{m_f^2}{M_{H_p}^2}\biggl[45
      +\biggl(\frac{194}{3}+\frac{16}{3}\log2\biggr)\zeta(2)+54\zeta(3)
          -N\biggl(\frac{22}{9}+4\zeta(2)+\frac{4}{3}\zeta(3)\biggr)\biggr]
                                                             \nonumber\\
 && \quad \hspace{20mm}
    -\biggl[\frac{87}{4}-\frac{13}{18}N
            -\frac{m_f^2}{M_{H_p}^2}\biggl(31-\frac{26}{9}N\biggr)\biggr]
                \log\frac{M_{H_p}^2}{m_f^2}
                                                             \nonumber\\
 && \quad \hspace{20mm}
    -\biggl[\frac{3}{4}-\frac{1}{6}N
       +\frac{m_f^2}{M_{H_p}^2}\biggl(5+\frac{2}{3}N\biggr)\biggr]
                 \log^2\frac{M_{H_p}^2}{m_f^2}
                                                             \nonumber\\
 && \quad \hspace{9mm}
    -\biggl(\frac{8}{3}-\frac{32}{3}\frac{m_f^2}{M_{H_p}^2}\biggr)
        \sum_{m_l<M_{H_p}}\Delta\biggl(\frac{m_l}{m_f}\biggr)
      +4\sum_{m_{v}<M_{H_p}}\frac{m_{v}^2}{M_{H_p}^2} \biggr>\biggr\}.
\label{eq:Polemassresf}
\end{eqnarray}
    The above result
confirms the asymptotic form (\ref{eq:2loopexpan}) of the two-loop
exact result (\ref{eq:2loop}), while the $O(\alpha_s^2)$ expression is new.

The numerical values of $\Delta(m_l/m_f)$,
defined in the eq.(\ref{eq:delta}), are given in the table 1 of the
ref.\cite{My}.
The quark masses are estimated from QCD sum rules:
$m_b=4.72$ GeV \cite{mb}, $m_c=1.46$ GeV \cite{mc}, $m_s=0.27$ GeV
\cite{ms}, $m_u+m_d \approx m_s/13 = 0.02$ GeV \cite{mud}.

For the decay mode $H_p\rightarrow b\overline{b}$
with five participating quark flavors one
obtains numerically:
\begin{eqnarray}
\lefteqn{\Gamma_{H_p\rightarrow b\overline{b}}
     =\frac{3\sqrt{2}G_FM_{H_p}}{8\pi}C_{H_{p}ff}(\beta)m_b^2
     \biggl\{\biggl(1-\frac{4m_b^2}{M_{H_p}^2}\biggr)^{\frac{1}{2}}
  +\frac{\alpha_s^{(5)}(M_{H_p})}
      {\pi}\delta^{(1)}\biggl(\frac{m_b^2}{M_{H_p}^2}\biggr)
       \biggl(1-\frac{4m_b^2}{M_{H_p}^2}\biggr)^{\frac{1}{2}} }
                                                             \nonumber\\
 && \quad
  +\biggl(\frac{\alpha_s^{(5)}(M_{H_p})}{\pi}\biggr)^2
   \biggl[-2.23039+169.22983\frac{m_b^2}{M_{H_p}^2}
                                                             \nonumber\\
 && \quad
     -\biggl(18.13889-16.55556\frac{m_b^2}{M_{H_p}^2}\biggr)
                        \log\frac{M_{H_p}^2}{m_b^2}
     +\biggl(0.08333-8.33333\frac{m_b^2}{M_{H_p}^2}\biggr)
                        \log^2\frac{M_{H_p}^2}{m_b^2}
                                                              \nonumber\\
  && \quad  \hspace{15mm}
          -\biggl(2.66667-10.66667\frac{m_b^2}{M_{H_p}^2}\biggr)
         \sum_{m_l\leq m_b}\Delta\biggl(\frac{m_l}{m_b}\biggr)
           +4\sum_{m_{v}\leq m_b}\frac{m_{v}^2}{M_{H_p}^2} \biggr]\biggr\},
\label{eq:Polemassresb}
\end{eqnarray}
where $\delta^{(1)}$ is defined in eq.(\ref{eq:2loop}).

    In ref.\cite{My} it was observed that the high order
corrections reduce the scale dependence significantly and resolve
the large discrepancy between the results for the decay rate of SM
Higgs in terms of running and pole quark masses. The theoretical
uncertainty was estimated at 5\%. The same effects were observed for
the decay rate of pseudoscalar Higgs particle. In the present calculation
one also estimates the theoretical uncertainty at approximately 5\%.

    It should be stressed, that the virtual top quark, which appears
in some topological types of three-loop diagrams (see fig.1 and fig.2 in
ref.\cite{My}) may give a nonnegligible contribution. In addition,
one may include a heavy virtual superpartners. However, the superpartners
decouple \cite{DecTheor} if the Higgs mass is much smaller than the
supersymmetry scale.

\vspace{5mm}

{\bf Acknowledgments}  It is a pleasure to thank D.Soper for
discussions, D.Broadhurst and B.Kniehl for helpful communications.
This work was supported by the
U.S. Department of Energy under grant No. DE-FG06-85ER-40224.

\end{document}